\documentstyle[preprint,aps]{revtex}
\begin{document}
\draft
\preprint{}
\title{
Effective Masses and Sizes of N(939), $\Delta\!\!\mbox{\boldmath(1232)}$
and N(1440) \\
in Nuclear Medium
}

\author{Hiroko Ichie\footnotemark[1], Arata Hayashigaki\footnotemark[2],
and Akira Suzuki}

\address{Department of Physics, Science University of Tokyo, Tokyo 162,
Japan}

\author{Masahiro Kimura}
\address{Suwa College, Science University of Tokyo, Nagano 391-02, Japan}

\footnotetext[1]{Present address: Department of Physics,
Tokyo Mtropolitan University, Tokyo 192-03, Japan}

\footnotetext[2]{Present address: Department of Physics,
Niigata University, Niigata 950-21, Japan}

\maketitle

\begin{abstract}

The effective masses and sizes of N(939), $\Delta(1232)$ and N(1440) in
nuclear medium are calculated for a model in which quarks are subjected
to a confinement force quenched by the scalar part of nuclear
interactions.
At the nuclear matter density, the nucleon swells by 50\% relative to
the free nucleon.
The masses of $\Delta(1232)$ and N(1440) decrease in nuclear medium
more than the nucleon, so that the pion requires less energy to excite
them in nuclear medium than it does for a free nucleon.

\end{abstract}

\clearpage
\narrowtext

The mass of the nucleon has its dynamical origin in the fundamental
dynamics of hadrons, quantum chromodynamics (QCD).
The most important ingredient of QCD for the nucleon mass is the
confinement force for three quarks in the nucleon.
When a nucleon is brought into nuclear medium, it is exposed to complex
interactions with other nucleons.
A part of the interaction which transforms as a Lorentz scalar can be
absorbed into the mass of the nucleon to define the effective mass.
In this paper we try to describe the change from the free nucleon mass
to the effective mass by quenching the confinement strength due to
the scalar part of the nuclear interaction.

As a consequence of this quenching, a nucleon swells in nuclear medium.
The notion of nucleon swelling has received much attention in connection
to the interpretation of the earlier EMC experiments,\cite{aub83}
and the Okamoto-Nolen-Schiffer anomaly.\cite{oka64}
The recent experiments of the low energy scattering of $K^{+}$ on some
nuclei suggest that the nucleon swelling actually takes place within
nuclei.\cite{mar90}
The simple picture provided by our model also enables us to discuss
the effective masses of other baryons such as $\Delta(1232)$ and N(1440)
in nuclear medium.
The purpose of this paper is thus twofold: to investigate both nucleon
swelling and the effective masses of $\Delta(1232)$ and N(1440) in
nuclear medium.

We use the bag model whose surface has a dynamical degree of freedom
with the radius $R$ and its conjugate momentum $P_{b}$.
The model was proposed by Brown, Durso and Johnson,\cite{bro83} and by
Tomio and Nogami (TN).\cite{tom85,suz86}
The Hamiltonian describing SU(2) baryons, of which we are interested in
N(939), $\Delta(1232)$ and N(1440) (denoted simply as N, $\Delta$ and R
hereafter), is given by
\begin{equation}
H = \sum_{i=1}^{3} \{ \mbox{\boldmath $\alpha$}_{i} \cdot
\mbox{\boldmath $p$}_{i} + \beta_{i} V(R,r_{i}) \} + \frac{g_{s}}{R}
\sum_{i>j} \mbox{\boldmath $\sigma$}_{i} \cdot
\mbox{\boldmath $\sigma$ }_{j} + H_{b} \;, \label{eq:h}
\end{equation}
where $V(R,r_{i})$ is zero for $r_{i}<R$ and infinity for $r_{i}>R$.
We added the approximate hyperfine interaction with the coupling
constant $g_{s}$ to the usual bag Hamiltonian to remove the degeneracy
between two spin-partners of the SU(2) baryons such as N and
$\Delta$.\cite{deg75}
The Hamiltonian to describe the dynamical motion of the bag is
\begin{equation}
H_{b} = \frac{P_{b}^{2}}{2M_{B}} + \frac{4\pi}{3}B_{D}R^{3} -
        \frac{Z_{0}}{R}+U_{\pi}(R) \label{eq:hb}
\end{equation}
with an $ad\;\;hoc$ inertial mass $M_{B}$ which is assumed to be
proportional to the mass of N or $\Delta$ with a proportional constant
$\beta$.\cite{bro83}
The term $U_{\pi}$ in Eq.~(\ref{eq:hb}) is the self-energy due to the
$\pi$-quark interaction, containing the intermediate states restricted
to N or $\Delta$.
We use the $U_{\pi}$ of TN with a little modification:
\begin{equation}
U_{\pi} = - 3\frac{f_{N}^{2}} {\pi m_{\pi}^{2}} \int_{0}^{\infty} dk
 \frac{k^{4}v^{2}(k,R)} {\omega_{k}(\omega_{k}+E_{N}-m_{N})}
- \frac{96}{25} \frac{f_{N}^{2}} {\pi m_{\pi}^{2}}
 \int_{0}^{\infty} dk \frac{k^{4}v^{2}(k,R)}
 {\omega_{k}(\omega_{k}+E_{\Delta}-m_{N})} \label{eq:upin}
\end{equation}
for N and R, and
\begin{equation}
U_{\pi} = - 3\frac{f_{N}^{2}}{\pi m_{\pi}^{2}} \int_{0}^{\infty} dk
 \frac{k^{4}v^{2}(k,R)}{\omega_{k}(\omega_{k}+E_{\Delta}-m_{\Delta})}
- \frac{24}{25} \frac{f_{N}^{2}}{\pi m_{\pi}^{2}} {\mbox{P}}
 \int_{0}^{\infty} dk
 \frac{k^{4}v^{2}(k,R)}{\omega_{k}(\omega_{k}+E_{N}-m_{\Delta})}
\label{eq:upid}
\end{equation}
for $\Delta$, where $f_{N}$ is the $\pi$N axial-vector coupling constant,
$m_{\pi}$ the pion mass, $\omega_{k}=(m_{\pi}^{2}+k^{2})^{1/2}$, and
$E_{B}=(m_{B}^{2}+k^{2})^{1/2}$ for B=N and $\Delta$.
The form factor is defined by
\begin{equation}
v(k,R) = \frac{3j_{1}(kR)}{kR} \;
\exp \left[-\frac{k^{2}R_{\pi}^{2}}{6} \right] \;,
\label{eq:ff}
\end{equation}
where $R_{\pi}$ is a phenomenological parameter related to pion radius.
Note that we ignore the isospin degree of freedom so that the proton and
the neutron are degenerate.

We apply the adiabatic approximation to separate the bag dynamics from
quark motion.\cite{tom85}
In this scheme one estimates the quark energy by ignoring the bag
motion; the quark energy thus obtained is a function of the bag radius
$R$ and is regarded as the potential for the bag motion.
Then N and $\Delta$, with $S=1/2$ and $3/2$, emerge as the states with
the lowest eigenvalues of the equation describing the bag motion:
\begin{equation}
\left[ H_{b} + \frac{3\eta_{0}}{R} \mp \frac{3g_{s}}{R} \right] b_{B}(R)
= m_{B}b_{B}(R) \;, \label{eq:hbvb}
\end{equation}
where the lower suffix $B$ stands for N or $\Delta$, $\eta_{0}=2.04$,
and the upper and lower signs correspond to N and $\Delta$,
respectively.
In this model the R is interpreted to be the first breathing excitation
of the nucleon.

The model contains six parameters of which $f_{N}^{2}$ is fixed to be
$0.08$.
We also use the MIT value for $B_{D}$:
$B_{D}^{1/4}=0.145$GeV.\cite{deg75}
The pion charge radius was measured to be around $0.66$fm.
However this should not be identified with the spatial extension of
quark components which we regard as $R_{\pi}$.
Several estimates put $R_{\pi}$ between $0.33$fm and
$0.40$fm.\cite{wei84}
We regard any $R_{\pi}$ in this interval as the experimental value of
the root-mean-square (rms) radius of the pion.
The rest of the parameters are determined to fit the observed masses of
N, $\Delta$ and R, and are $\beta=0.516$, $Z_{0}=2.201$ and
$g_{s}=0.226$, which yield $R_{\pi}=0.38$fm.

Let us turn to baryons in nuclear medium.
We attempt to replace complex nuclear interactions on a baryon by
the quenched confinement force for three quarks in the baryon.
It is specified in such a way that Eq.~(\ref{eq:hbvb}) for the nucleon
gives the effective mass obtained from conventional nuclear physics.
We use the standard $\sigma\omega$ model, referring to \cite{sw86} for
the details.
The effective mass of the nucleon, $m_{N}^{*}$, is defined by
\begin{equation}
m_{N}^{*} = m_{N} - g_{\sigma}\sigma  \label{eq:mnstar}
\end{equation}
in the mean-field approximation for $\sigma$.
Quantities appearing with an asterisk hereafter are to be understood as
the quantities in nuclear medium.
The mean-field $\sigma$ satisfies the following equations:
\begin{equation}
g_{\sigma}\sigma = \frac{g_{\sigma}^{2}}{m_{\sigma}^{2}}[\rho_{S}
-b_{\sigma}m_{N}(g_{\sigma}\sigma)^{2}
-c_{\sigma}(g_{\sigma}\sigma)^{3}] \;, \label{eq:gsig}
\end{equation}
\begin{equation}
\rho_{S} = \frac{4}{(2\pi)^{3}}\int d^{3}k
\frac{m_{N}^{*}}{\sqrt{k^{2}+m_{N}^{*}{}^{2}}}\theta(k_{F}-k) \;,
\label{eq:rhos}
\end{equation}
where we consider our system as nuclear matter filled by nucleons up to
the level of $k_{F}$.
We use $g_{\sigma}^{2}=154.050$, $b_{\sigma}=0.000673$,
$c_{\sigma}=0.009786$ and $m_{\sigma}=550$MeV, which were obtained by
Gmuca to reproduce the Dirac-Bruckner-Hartree-Fock results of nuclear
matter calculation.\cite{gmu92}
Two typical values of the effective mass parameter defined by
$\alpha_{N}=m_{N}^{*}/m_{N}$ are 0.691 and 0.619 for $k_{F}=1.20$ and
$1.36\;\rm{fm}^{-1}$, respectively.
The value of $k_{F}=1.36\;\rm{fm}^{-1}$ is for the normal nuclear
matter and $1.20\;\rm{fm}^{-1}$ is to simulate medium heavy nuclei.

Now we regard the change of $m_{B}$ to $m_{B}^{*}$ as the consequence
of quenching of the parameters in our quark model due to nuclear
interactions.
We keep $\beta$, $Z_{0}$ and $g_{s}$, which are related to the intrinsic
nature of the bag interior, constant throughout.
Thus $f_{N}$, $m_{\pi}$, $R_{\pi}$ and $B_{D}$ are left as the
parameters to be quenched by nuclear interactions.

The effect of nuclear interactions on $f_{N}$ was studied by Rho under
the assumption of the relation $g_{N}/g_{A}=g_{N}^{*}/g_{A}^{*}$, where
$g_{N}$ and $g_{A}$ are the $\pi$N pseudo-scalar and the axial-vector
coupling constants, respectively.\cite{rho85}
The equivalence theorem connecting $g_{N}$ to $f_{N}$ derives
\begin{equation}
\frac{(f_{N}/m_{\pi})^{*}}{(f_{N}/m_{\pi})}
= \frac{1}{\alpha_{N}}\frac{g_{A}^{*}}{g_{A}} \;. \label{eq:fmfm}
\end{equation}
We quote \cite{rho85} for the calculation of $g_{A}^{*}$.

We know the properties of pions in nuclear medium as little as those
of the free pion.
The particular properties of pions are associated with the chiral
symmetry of the strong interaction and are beyond the simple bag-based
description.
Let us use the following conventional model to describe the pion.
The effective Hamiltonian is given by
\begin{equation}
H_{M} = \sum_{i=1}^{2} \{ \mbox{\boldmath $\alpha$}_{i} \cdot
\mbox{\boldmath $p$}_{i} + \beta_{i} V(R,r_{i}) \} + 2 \frac{g_{s}}{R}
\mbox{\boldmath $\sigma$}_{1} \cdot \mbox{\boldmath $\sigma$}_{2}
+ \frac{4\pi}{3}B_{D}R^{3} - \frac{Z_{\pi}}{R}
       + H_{q\bar{q}} \;, \label{eq:hm}
\end{equation}
where $Z_{\pi}$ is an adjustable parameter corresponding to $Z_{0}$ for
baryons.
We introduced the phenomenological $\mbox{q}\bar{\mbox{q}}$ interaction,
$H_{q\bar{q}}$.
It is needed to bring $m_{\pi}$ down to the observed value, otherwise
$m_{\pi}$ comes out to be too large.
Presumably q and $\bar{\mbox{q}}$ will interact strongly when they are
at the largest separation inside the bag.
Considering this, we take $H_{q\bar{q}}$ to be\cite{bww83}
\begin{equation}
H_{q\bar{q}} = - \lambda^{2} R^{3}\delta (r_{1}-R)
\delta^{(3)}(\mbox{\boldmath $r$}_{1}+\mbox{\boldmath $r$}_{2}) \;.
\label{eq:hqq}
\end{equation}
We ignored the bag motion here for simplicity so that the pion emerges
as the ground state with the lowest eigenvalue minimized with respect
to $R$.

In order to solve the eigenvalue equation for $H_{M}$, we diagonalize
it within the basis of the eigenstates of the Hamiltonian without
$H_{q\bar{q}}$.
Since we are interested in the ground state, it will be sufficient to
truncate the diagonalization space with the first three states
($1\mbox{S}_{1/2}$, $1\mbox{P}_{3/2}$, $1\mbox{P}_{1/2}$) which
accommodate the pair of q and $\bar{\mbox{q}}$.
Two parameters, $Z_{\pi}=2.257$ and $\lambda=0.256$, were determined by
fitting $m_{\pi}=138\mbox{MeV}$ and $R_{\pi}=0.38$fm.
We assume that these parameters are out of the influence of nuclear
interactions for the same reason as in the baryon case.
Hence only the quenched $B_{D}$ determines $m_{\pi}^{*}$ and
$R_{\pi}^{*}$ in nuclear medium.
The quenched $B_{D}$ for the pion may be different from that for the
nucleon.
However, our inadequate knowledge of pion properties in nuclear medium
leads us to a common $B_{D}^{*}$ for both the nucleon and pion.

We determine $B_{D}^{*}$ according to the following procedure.
The Hamiltonian describing baryons in nuclear medium depends on
$m_{N}^{*}$, $m_{\Delta}^{*}$ and $B_{D}^{*}$ through $f_{N}^{*}$,
$M_{B}^{*}$ and $U_{\pi}$ in which we replace
$1/(\omega_{k}+E_{N}-m_{B})$ by
$\theta(k-k_{F})/(\omega_{k}^{*}+E_{N}^{*}-m_{B}^{*})$
with B=N in Eq.~(\ref{eq:upin}) and with B=$\Delta$ in
Eq.~(\ref{eq:upid}) to take account of the Pauli principle.
The lowest eigenvalues of Eq.~(\ref{eq:hbvb}) are $m_{N}^{*}$ for
$S=1/2$ and $m_{\Delta}^{*}$ for $S=3/2$.
Thus the eigenvalue equation is a set of coupled-selfconsistent
equations for $m_{N}^{*}$ and $m_{\Delta}^{*}$, and hence for
$\alpha_{N}$ and $\alpha_{\Delta}$.
One can determine $B_{D}^{*}$ such that the selfconsistency is satisfied
for a given $\alpha_{N}$.
Once $B_{D}^{*}$ is fixed, the effective masses of excited baryons
can be evaluated by solving the eigenvalue equation (\ref{eq:hbvb}).
The rms radius of a baryon B is calculated according to
\begin{equation}
\langle r^{2} \rangle _{B(q)}^{1/2} = 0.73 \;
\left[ \int_{0}^{\infty} dR b_{B}^{*}(R) R^{2} b_{B}(R) \right]^{1/2} \;.
\label{eq:rms}
\end{equation}

The effective masses and the nucleon rms radius thus obtained are shown
in Figs.1 and 2, respectively.
Obviously the nucleon swells in nuclear medium as nucleon density is
increased and the rms radius reaches
$[\langle r^{2} \rangle_{N(q)}^{1/2}]^{*}/
[\langle r^{2} \rangle_{N(q)}^{1/2}]=1.96$ at the nuclear matter density.
This large number is not particularly surprising.
The rms radius evaluated by Eq.~(\ref{eq:rms}) is of the quark core.
Consider that the physical nucleon consists of the three-quark core and
a pion.
Then the rms radius is given by\cite{suz86}
\begin{equation}
\langle r^{2} \rangle_{N} =
P_{3q} \langle r^{2} \rangle_{N(q)} +
(1-P_{3q}) \langle r^{2} \rangle_{N(\pi)} \label{eq:r2n} \; ,
\end{equation}
where $P_{3q}$ is the probability to find the nucleon in the three-quark
state and is given by
\begin{eqnarray}
\lefteqn{P_{3q}} \nonumber \\
& = & \left[1+\frac{3f_{N}^{2}} {\pi m_{\pi}^{2}} \int_{0}^{\infty} dk
\frac{k^{4}\bar{v^{2}}(k)} {\omega_{k}(\omega_{k}+E_{N}-m_{N})^{2}}
 + \frac{96f_{N}^{2}}{25\pi m_{\pi}^{2}}
 \int_{0}^{\infty} dk \frac{k^{4}\bar{v^{2}}(k)}
 {\omega_{k}(\omega_{k}+E_{\Delta}-m_{N})^{2}}\right]^{-1} \label{eq:p3q1}
\nonumber \\
 & &
\end{eqnarray}
with
$\bar{v^{2}}(k)=\int_{0}^{\infty} dR b_{N}^{*}(R)v^{2}(k,R)b_{N}(R)$.
The second term in Eq.~(\ref{eq:r2n}) is the pion contribution.
It should be understood, when we calculate $P_{3q}$ in nuclear medium,
we replace the quantities in Eq.~(\ref{eq:p3q1}) by the corresponding
ones with an asterisk and make the similar replacement made in $U_{\pi}$
to account for the Pauli principle.
Our calculation gives $P_{3q}=0.74$, 0.87 and 0.91 for $k_{F}=0.00$,
1.20 and $1.36\rm{fm}^{-1}$, respectively.
The measured value for a free nucleon,
$\langle r^{2}\rangle_{N}^{\rm{exp}} = (0.86\;\rm{fm})^{2}$, gives
$\langle r^{2} \rangle_{N(\pi)} = (1.27\;\rm{fm})^{2}$.
Let us use this value in nuclear medium, anticipating that the
pionic contribution becomes less significant as the density increases.
The result is shown by the dashed line in Fig.2.
The typical values are $\langle r^{2} \rangle_{N}^{1/2}=1.13$ and
$1.29\rm{fm}$ for $k_{F}=1.20$ and $1.36\rm{fm}^{-1}$ respectively.
This tells us that the nucleon radius swells by 50\% at the nuclear
matter density.
This number is close to the 45\% obtained in the skyrmion
picture.\cite{rho85}
Can one estimate the nucleon density $n_{c}$ at which a swollen bag
comes into contact with another so that quarks in the bag may squeeze
into the other?
Let us define $n_{c}$ such that the volume of the quark core swells to
occupy the mean volume shared by a nucleon at $n_{c}$, hence
$\langle r^{2} \rangle_{N(q)}^{1/2}=(9\pi/8)^{1/3}/k_{F}$ at $n_{c}$.
This condition is reached at $k_{F}=1.28\rm{fm}^{-1}$ which corresponds
to 0.8 times the normal nuclear matter density.

In regard to the effective masses, $\Delta$ reduces its mass more
than the nucleon.
Consequently the pion requires less energy to excite the nucleon to
the $\Delta$ in nuclear medium than it does outside.
The mass difference between $\Delta$ and N decreases to 50\% of that
for free particles at the nuclear matter density.
Our number is much larger than the 20\% predicted in the
Nambu-Jona-Lasinio model.\cite{chr90}
In our definition of the effective mass only the interaction which
transforms as a Lorentz scalar is taken into account, so that the
nucleons with the effective mass are still interacting with each other
through the vector field.
The vector interaction may work in a different way for N and $\Delta$,
and must be taken into account for quantitative discussion of the
downward shift of the $\Delta$ resonance observed in $\pi$-nucleus
scattering.\cite{ost92}
Our result also predicts that R will be excited with less energy in
$\pi$-nucleus scattering than in $\pi$N scattering.
This is signalled by the large inelasticity in the two-pion
channel at lower pion energy than in $\pi$N scattering.

We would like to thank Professor Yuki Nogami and Professor Rajat K.
Bhaduri for their helpful discussions and valuable comments.

\clearpage

\vspace{3cm}
\begin{center}
Figure Captions \\
\end{center}
\vspace{1cm}
\noindent
Fig.1

The effective masses of N(939) (the solid line), $\Delta(1232)$ (the
dashed line) and N(1440) (the dot-dashed line).
Also the mass difference, $\delta m=m_{\Delta}^{*}-m_{N}^{*}$, is shown
by the dotted line.
The arrow indicates the Fermi momentum corresponding to the nuclear
matter density. \\

\vspace{1cm}
\noindent
Fig.2

The root-mean-square radii of the nucleon in nuclear medium.
The solid line represents the rms radii of the three-quark core in
the nucleon, and the dashed line those of the physical nucleon estimated
by using Eq.~(\ref{eq:r2n}).
The arrow indicates the Fermi momentum corresponding to the nuclear
matter density. \\

\end{document}